# What is Large in Large-Scale?
# *A Taxonomy of Scale for Agile Software Development*[1]

Torgeir Dingsøyr,[1,2] Tor Erlend Fægri,[1] Juha Itkonen[3]

[1]SINTEF,
NO-7465 Trondheim, Norway
torgeird@sintef.no, toref@sintef.no

[2]Department of Computer and Information Science,
Norwegian University of Science and Technology

[3]Aalto University, Department of Computer Science and Engineering
FI-00076 Aalto, Finland
juha.itkonen@aalto.fi

**Abstract.** *Positive experience of agile development methods in smaller projects has created interest in the applicability of such methods in larger scale projects. However, there is a lack of conceptual clarity regarding what large-scale agile software development is. This inhibits effective collaboration and progress in the research area. In this paper, we suggest a taxonomy of scale for agile software development projects that has the potential to clarify what topics researchers are studying and ease discussion of research priorities.*

---

# 1 Introduction

In the introduction to the special issue on agile methods in IEEE Computer in 2003, Williams and Cockburn stated that "agile value set and practices best suit co-located teams of about 50 people or fewer who have easy access to user and business experts and are developing projects that are not life-critical" [19]. Since then, agile methods have received significant attention from practitioners and academia [6], and have increasingly been applied in new settings, such as global, distributed development [16] and large-scale development [13].

The rise of agile methods has also brought out critics, e.g. related to lack of focus on architectural decisions and that the methods are suitable only for small teams. Large projects are likely to have high societal impact and thus justify a serious consideration of critique. It is very important to understand if, when, and how agile methods can be suitable 'in the large.' Adaptions of agile principles may be necessary when scaling along dimensions of project size, project complexity and distribution of personnel.

The continued interest in this research area is exemplified by the practitioners at the XP2010 conference voting "agile in the large" to be "the top burning research question" [8]. Boehm and Turner discussed how risk exposure can be used to balance agile and plan-driven methods [2]. Lindvall et al. reported from meetings and a workshop amongst large companies and their experience with agile in the large [10].

However, there is little agreement on what large-scale agile development is [5]. Webster´s define 'large-scale' as "very extensive; of great scope" [18]. Some have used the term to describe projects with many members in a single team, while others are referring to projects with multiple teams over a number of years or a combination of size, distribution and specialization [3].

In order to facilitate discussion of planned studies and identify basic assumptions and knowledge gaps, we suggest a taxonomy of scale for agile projects, and discuss how this taxonomy can be used in future studies of large-scale agile development.

Research on large-scale agile development can include numerous topics, but the research community should emphasize more conceptual clarity and awareness regarding scale and the implications for scalability of agile methods. This would, for example, contribute to effective selection of case studies. To deepen our understanding and develop research-based knowledge, we need in-depth studies that serve as exemplars [7]; they provide a richer description of the projects and help research to connect to relevant theories that can explain the cases and thereby provide lessons for other projects. An agreement upon a taxonomy of scale for agile development projects would make it clearer what topics researchers are addressing, ease discussion of priority of topics in research agendas and make it more evident when studies can provide meaningful lessons for others.

## 2 A taxonomy of scale

A taxonomy may pinpoint differences in various types of large-scale agile projects that could lead to novel research questions. If we are to develop such a taxonomy of scale, the question is then which dimension(s) should we use? Project cost, number of people involved, number of requirements, lines of code, number of teams, additional practices needed? Would large-scale be different in various application domains [14]? In the following, we discuss these possible dimensions:

When focusing on large-scale in relation to development method, the cost is not a sufficient a criterion for large-scale. Costs vary across projects – some may involve hardware procurement or organizational change programs. Furthermore, these cost drivers are different from country to country. The code size is also a problematic factor; code could be generated by tools or be the result of modifications to existing code. Number of requirements, user stories or features to be developed suffers from high variability in the time to implement them. Some domains have a number of non-functional requirements, such as real-time systems in the telecom industry, which leads to additional effort in development. In addition, the size of software in terms of code or requirements is rarely available as comparable measures across technologies and project contexts.

We suggest including generally available and reliable factors. One factor that makes large projects difficult is the coordination overhead that is increasing with size. In management science, there are two general approaches to coordination of work: programming (up-front decisions) and feedback [11]. The nature of software development, being innovative work that is only partially compatible with programmed coordination, requires a strong emphasis on personal communication [9] and tacit knowledge embedded in the team [1]. The common advice in agile methods is to have teams of 7 plus/minus two people in a team[2] to achieve effective teamwork by reducing the number of communication lines.

When more people are required in a project, the work is then divided between several teams. We can identify this as a second type of large-scale project. The use of multiple teams will reduce the effectiveness of communication [17]. Curtis [4] shows how rapid clarification of conflicts is essential to effective software practice. More teams will incur an increase in the number of communication lines, and we can identify a new major change to coordination when we exceed 7 (+/- 2) teams. Coordination forums with many participants will be ineffective, and therefore large projects needs additional fora to coordinate subprojects.

---
[2]http://www.scrum.org/Portals/0/Documents/Scrum%20Guides/Scrum_Guide.pdf.

This third type of large projects needs a new level of coordination. A pyramid organization paradigm adds distance between floor and top [15], and distance increases risk of distortion in information and 'knowledge silos'.

Following this line of thought, we end up with a taxonomy as described in Table 1. This is inspired by a taxonomy in requirements engineering [12]. One could criticize that this taxonomy is based on a theoretical model of a project, and in practice one may organize a large project with subprojects that are functionally or technically divided. But distinguishing on number of teams makes an easy and widely applicable taxonomy.

**Table 1.** A taxonomy of scale of agile software development projects.

| *Level* | *Number of teams* | *Coordination approaches* |
|---|---|---|
| Small-scale | 1 | Coordinating the team can be done using agile practices such as daily meetings, common planning, review and retrospective meetings. |
| Large-scale | 2-9 | Coordination of teams can be achieved in a new forum such as a Scrum of Scrums forum. |
| Very large-scale | 10+ | Several forums are needed for coordination, such as multiple Scrum of Scrums. |

We suggest that this taxonomy can be used in designing studies in order to be more precise on selection criteria in case studies. Further, the taxonomy could be used in research question design, in order to focus on relations between large-scale projects and topics such as appropriateness of agile practices and when additional practices are required. Finally, the taxonomy can be important in characterizing state of the art of research, in showing the state of research on the different levels of scale. We would welcome a further discussion on the suitability of this taxonomy and whether a taxonomy of scale should include also other dimensions.

**Acknowledgement:** The work on this article was supported by the SINTEF internal project "Agile project management in large development projects" and by the project Agile 2.0 which is supported by the Research council of Norway through grant 236759/O30, and by the companies Kantega, Kongsberg Defence & Aerospace and Steria.